\documentclass[a4paper]{jpconf}
\usepackage{graphicx}
\newcommand{\ba}{\begin{eqnarray}}
\newcommand{\ea}{\end{eqnarray}}
\newcommand{\beqs}{\begin{eqnarray}}
\newcommand{\eeqs}{\end{eqnarray}}

\begin{document}
\title{MOMENTUM TRANSFER DEPENDENCE OF GPDs}

\author{O.V. Selyugin}

\address{BLTPh, Joint Institute for Nuclear Research, Dubna, Russia  }

\ead{selugin@theor.jinr.ru}

\begin{abstract}
 Based on the factorization representation of the
 General Parton Distributions (GPDs) the momentum transfer dependence
 was determined by the analysis of the different representations of parton distribution functions
  (PDFs) and all possible experimental data of the electromagnetic form factors of the proton and neutron.
  The obtained $t$-dependence of the GPDs is checked by analysis of the different hadronic
  reactions (including exclusive and elastic hadron scattering) in a wide energy region
   with minimum free fitting parameters.
\end{abstract}
\vspace{7.2mm}
\section{Introduction}

   The parton picture of the hadron  structure
   is represented, in most part, by the parton distribution functions (PDFs).
   They are determined in the deep inelastic processes. The next step in the development of the picture of the hadron
 structure
 was made by introducing the non-forward structure functions - general parton distributions - GPDs \cite{Mil94,Ji97,R97}
     with the spin-independent  $H(x,\xi,t) $
   and the spin-dependent $E(x,\xi,t)$ parts.
  Some of the advantages of  GPDs were presented by the sum rules \cite{Ji97}
   %which put the connections of the  GPDs with the standard electromagnetic hadron form factors.
  \ba
 F_{1}^q (t) = \int^{1}_{0} \ dx  \ {\cal{ H}}^{q} (x,\xi=0, t),  \ \ \
%\ea
%\ba
 F_{2}^q (t) = \int^{1}_{0} \ dx \  {\cal{E}}^{q} (x, \xi=0, t).
\ea
     Using the different  momenta of GPDs as a function of $x^{n-1}$  we can obtain the different form factors.
  \ba
 F_{n-1}^q (t) = \int^{1}_{0} \ dx  \ x^{n-1} {\cal{F}}^{q} (x,\xi=0, t)
\ea
  $n=0$, (zero moment)  give the Compton form factors $R_{V}(t), R_{A}(t), R_{T}(t)$ \\
 $n=1$, (first moment) give  the electromagnetic form factors $F_{1}(t), F_{2}(t)$ \\
 $n=2$, (second moment) give the gravimagnetic form factors $A_{1}(t), B_{2}(t)$ \\

  Good knowledge of the momentum transfer dependence of the GPDs  is required.
Now we cannot obtain the $t$-dependence of GPDs  from the first principles,
but it can be obtained from the phenomenological description by $GPDs$
of the nucleon electromagnetic form factors. Many different forms
  of the $t$-dependence of GPDs were proposed.
   In the quark diquark model \cite{Liuti1,Liuti2} the form of  GPDs
   consists of three parts - PDFs, function distribution and Regge-like.
 In other works (see e.g. \cite{Kroll04}),
  the description of the $t$-dependence of  GPDs  was developed
  in a  more complicated picture using the polynomial forms with respect to $x$.
    In \cite{Yuan03},  %,Burk04}
    it was shown that at  large $x  \rightarrow 1$
    and momentum transfer the behavior of GPDs
  requires a larger power of $(1-x)^{n}$ in the  $t$-dependent exponent:
\ba
{\cal{H}}^{q} (x,t) \  \sim  exp [ a \ (1-x)^n \ t ] \ q(x).
\ea
with $n \geq 2$. It was noted that $n=2$ naturally leads to the Drell-Yan-West duality
 between parton distributions at large $x$ and  form factors.

 \section{New momentum transfer dependence of GPDs }

Let us modify the original Gaussian ansatz
% in order to incorporate
% the observations of \cite{R98} and \cite{Burk04}
 and choose  the $t$-dependence of  GPDs in a simple form
\ba
{\cal{H}}^{q} (x,t) \  = q(x) \   exp [  a_{+}  \
    (1-x)^2/x^{m}  \ t ].
    \label{GPD0}
%\frac{(1-x)^2}{x^{m} } \ t ].
\ea
  The value of the parameter $m=0.4$ is fixed by the low $t$  experimental data while
 the free parameters $a_{\pm}$ ($a_{+} $ - for ${\cal{H}}$
and $a_{-} $ - for ${\cal{E}}$) were chosen to reproduce the
experimental data in the whole $t$ region.
  The isotopic invariance can be used to relate the proton and neutron GPDs.
 Hence, we do not change any parameter
 and keep the same $t$-dependence of GPDs as in the case of  proton.

 In all calculations we restrict ourselves %, as in other quoted works,
  to  the contributions of only valence  $u$ and $d$ quarks.
  In our first  work \cite{ST-PRDGPD} the function $q(x)$
 is based on the MRST02 global fit \cite{MRST02}.

%\section{GPDs and form factors of the nucleon }

  Further development of the model requires a careful analysis of the momentum transfer form of GPDs and
   properly chosen  PDFs form. In \cite{GPD-PRD14}, the analysis of  more than 24 different
   PDFs was made.
   The complex analysis of the corresponding description of the electromagnetic form factors of the proton and neutron
    by the different  PDF sets  (24 cases) was carried out . These
   PDFs include the  leading order (LO), next leading order (NLO) and next-next leading order (NNLO)
   determination of the parton distribution functions. They used the different forms of the $x$ dependence of  PDFs. % eqs. (\ref{sq1})-(\ref{ex3}).
    We slightly complicated the form of GPDs  in comparison with eq.(\ref{GPD0}),
   but it is the simplest one as compared to other works (for example \cite{DK-13}).
\ba
{\cal{H}}^{u} (x,t) \  = q(x)^{u}_{nf} \   e^{2 a_{H}  \ \frac{(1-x)^{2+\epsilon_{u}}}{(x_{0}+x)^{m}}  \ t };  \ \ \ % \\ \nonumber
{\cal{H_d}}^{d} (x,t) \  = q(x)^{d}_{nf} \   e^{2 a_{H} (1+\epsilon_{0}) (\frac{(1-x)^{1+\epsilon_{d}}}{(x_{0}+x)^{m}} ) \ t }.
\label{t-GPDs-H}
\ea
\ba
{\cal{E}}^{u} (x,t) \  = q(x)^{u}_{fl} \   e^{2 a_{E}  \ \frac{(1-x)^{2+\epsilon_{u}}}{(x_{0}+x)^{m}}  \ t }; \ \ \  % \\ \nonumber
{\cal{E_d}}^{d} (x,t) \  = q(x)^{d}_{fl} \   e^{2 a_{E}(1+\epsilon_{0}) (\frac{(1-x)^{1+\epsilon_{d}}}{(x_{0}+x)^{m}} ) \ t }.
\label{t-GPDs-E}
\ea
 where $q(x)^{u,d}_{fl}=q(x)^{u,d}_{nf} (1.-x)^{z_{1},z_{2}}$

 To obtain the form factors,  we have to integrate  over $x$ in the whole range $0 - 1$.
 Hence, the form of the $x$-dependence of PDF affects the form and size of the
  form factor.
 But the PDF sets are determined from the inelastic processes only in  some region of $x$, which is only
 approximated to $x=0$ and $x=1$.
   Some  PDFs  have the polynomial form of $x$ with
     different power.  Some other have the exponential dependence of $x$.
  As a result, the behavior of  PDFs, when $x \rightarrow 0$ or $x \rightarrow 1$,  can  impact  the
    form of the calculated form factors.
   The analysis was carried out with different forms of the
   $t$ dependence of GPDs. The minimum number of  free parameters was six and maximum was ten.
    The obtained electromagnetic form factors and
      the ratio of the electromagnetic form  factors
    for the proton $\mu_{p} G^{p}_{E}/G^{p}_{M}$ and
   for the neutron $\mu_{n} G^{n}_{E}/G^{n}_{M}$ were
    described quantitatively well.
      Our calculation reproduces the data obtained by the polarization method.

    On the basis of our GPDs with
    ABM12 \cite{ABM12} PDFs    we calculated the hadron form factors
     by the numerical integration
 \ba
 F_{1}(t) = \int^{1}_{0}  dx
%   \\	\nonumber
 [\frac{2}{3}q_{u}(x)e^{2 \alpha_{H} t (1.-x)^{2+\epsilon_{u}}/(x_{0}+x)^m} % \\	\nonumber
  -\frac{1}{3} q_{d}(x)e^{ 2 \alpha_{H}  t (1.-x)^{1+\epsilon_{d}}/((x_{0}+x)^{m})} ]
\ea
   and then
    by fitting these integral results by the standard dipole form with some additional parameters
    for  $F_{1}(t)$
      \begin{eqnarray}
   F_{1}(t)  = \frac{4m_p - \mu t }{4m_p -  t } \frac{1}{(1 + q/a_{1}+q^{2}/a_{2}^2 +  q^3/a_{3}^3)^2 }
 \label{Gt}
 \end{eqnarray}

  The matter form factor %$A(t)$
 \ba
  A(t)=  \int^{1}_{0} x \ dx
% \\	\nonumber
 [ q_{u}(x)e^{2 \alpha_{H} t (1.-x)^{2+\epsilon_{u}}/(x_{0}+x)^m}  % \\	\nonumber
  + q_{d}(x)e^{ 2 \alpha_{H}  t (1.-x)^{1+\epsilon_{d}}/((x_{0}+x)^{m})} ]
\ea
 is fitted   by the simple dipole form  $  A(t)  =  \Lambda^4/(\Lambda^2 -t)^2 $.

        Our description is valid up to a large momentum transfer
        with the following parameters:
        $a_{1}=16.7$ GeV,  \ $a_{2}^{2}=0.78$ GeV$^{2}$, \ $a_{3}^{3}=12.5$ GeV$^{3}$ and $\Lambda^2=1.6$ GeV$^{2}$.
        These form factors will be used in our model of the proton-proton and proton-antiproton elastic scattering.

\section{The Compton cross sections}

   Our calculations are based on the
    works \cite{R98,DK-13}.
    The differential cross section for that reaction can be written as
  \ba
  \frac{d\sigma}{dt} =  \frac{\pi \alpha^{2}_{em}}{s^{2}} \frac{(s-u)^{2}}{-u s}
  [R_{V}^{2}(t) \ - \ \frac{t}{4 m^{2}} R^{2}_{T}(t) % \\ \nonumber
    + \frac{t^{2}}{(s-u)^{2}} R^{2}_{A} (t)],
    \label{RCS}
\ea
  where $R_{V}((t)$, $R_{T}(t)$, $R_{A}(t)$ are the form factors given by the $1/x$
  moments of the corresponding GPDs $H^{q}(x,t)$,  $E^{q}(x,t)$, $\tilde{H}^{q}(x,t)$ .
     The last is related with the axial form factors.
     As noted in \cite{DK-13}, this factorization,
  which bears some similarity to the handbag factorization of DVCS,
  is formulated in a symmetric frame where the skewness $\xi=0$.
   For  $H^{q}(x,t)$,  $E^{q}(x,t)$ we used the PDFs obtained from the
   works \cite{Kh12} with the parameters
 %   the additional  parameters for  $E^{q}(x,t)$ which were
   obtained in our fitting procedure of the description of the proton and neutron electromagnetic
   form factors   in \cite{GPD-PRD14}.

  \ba
 R_{V}(t) % &=&  \sum_{q} e^{2}_{q} \int_{0}^{1} \frac{dx}{x} H_{q}(x,,\xi=0,t) \\ \nonumber
 =
  \int_{0}^{1} \frac{dx}{x} \left\{  \frac{4}{9} u(x) \ exp[2\alpha_{1} \frac{(1-x)^{p_{1}}}{(x_{0}+x)^{p_[2}} t]
   \ + \ \frac{1}{9} d(x) \ exp[2\alpha_{1} \frac{(1-x)^{p_{1}(k_{d})}}{(x_{0}+x)^{p_{2}}} t]  \right\} ,
\ea

  \ba
 R_{T}(t) = % &=&  \sum_{q} e^{2}_{q} \int_{0}^{1} \frac{dx}{x} E_{q}(x,,\xi=0,t) \\ \nonumber
 =
  \int_{0}^{1} \frac{dx}{x} \left\{  \frac{4}{9} u^{e}(x) \ exp[2\alpha_{1} \frac{(1-x)^{p_{1}}}{(x_{0}+x)^{p_[2}} t]
   \ + \ \frac{1}{9} d^{e}(x) \ exp[2\alpha_{1} \frac{(1-x)^{p_{1}(k_{d})}}{(x_{0}+x)^{p_{2}}} t]  \right\} ,
\ea

  \ba
 R_{A}(t) % &=&  \sum_{q} e^{2}_{q} \int_{0}^{1} \frac{dx}{x} \tilde{H}_{q}(x,,\xi=0,t) \\ \nonumber
 =
  \int_{0}^{1} \frac{dx}{x} \left\{  \frac{4}{9} \Delta u^{e}(x) \ exp[2\alpha_{1} \frac{(1-x)^{p_{1}}}{(x_{0}+x)^{p_[2}} t]
   \ + \ \frac{1}{9} \Delta d^{e}(x) \ exp[2\alpha_{1} \frac{(1-x)^{p_{1}(k_{d})}}{(x_{0}+x)^{p_{2}}} t]  \right\} ,
\ea
  For $\tilde{H}^{q}(x,t)$ we take $\Delta q^{e}$ in the form
   \ba
    \Delta q^{e} = N_i x^a_1 (1+ a_2 \sqrt{x} +a_3 x),
   \ea
   with the parameters determined in \cite{80-Flo-09}.
     The calculations of $R_{i}$  on the whole, correspond to the calculations \cite{DK-13},
     but the  integrals with our  $t$ dependence of GPDs do not divergence at momentum transfer $ -t > 2$ GeV$^2$.
     In \cite{DK-13},  $R_{i}$ is presented beginning with $-t=4$ GeV$^2$.

%%%Fig 14
\begin{figure}
%\begin{center}
%\mbox{\epsfysize=60mm\epsffile{GedGd.eps}}x
\includegraphics[width=.3\textwidth]{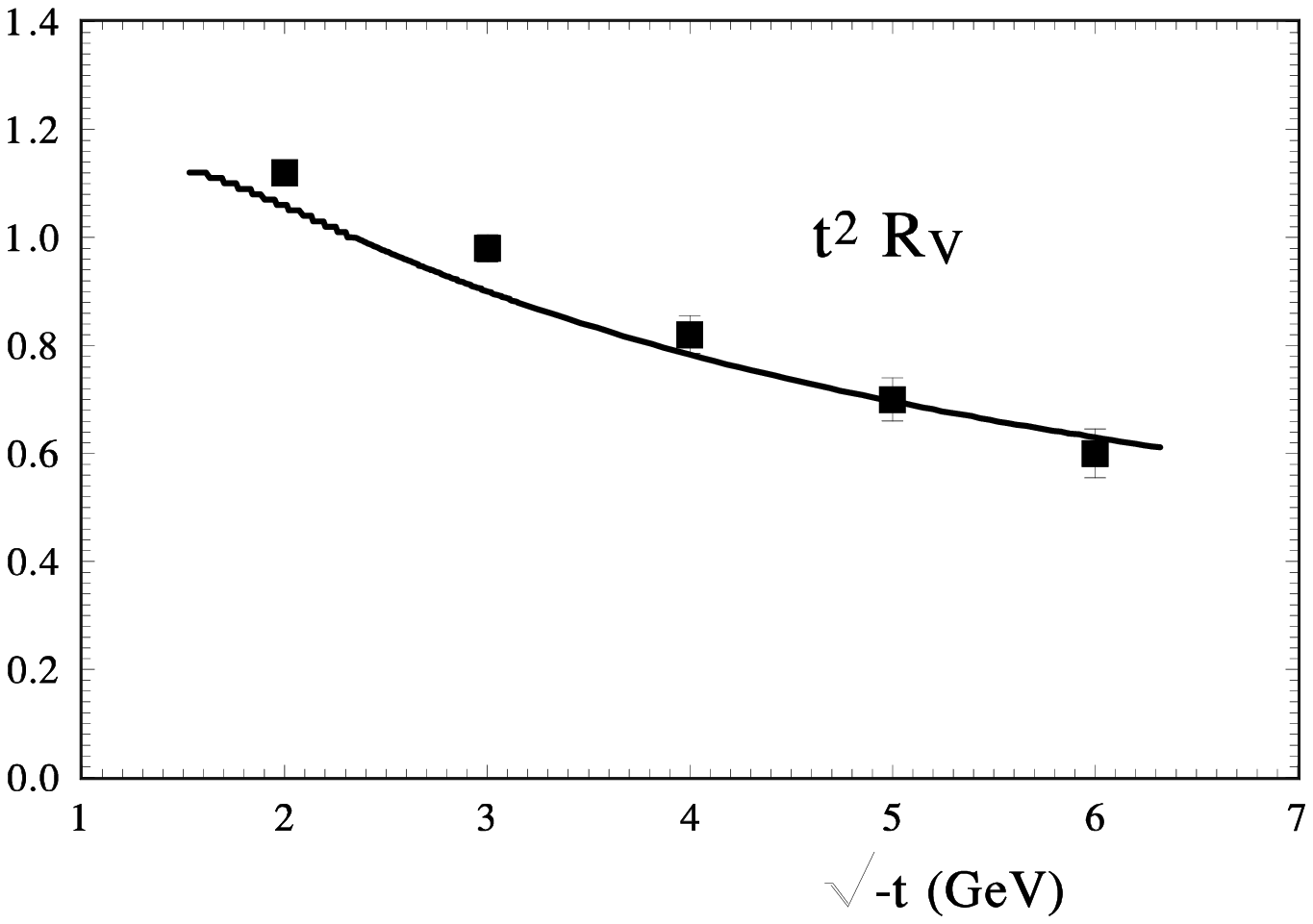} %{ffdm.ps}
\includegraphics[width=.3\textwidth]{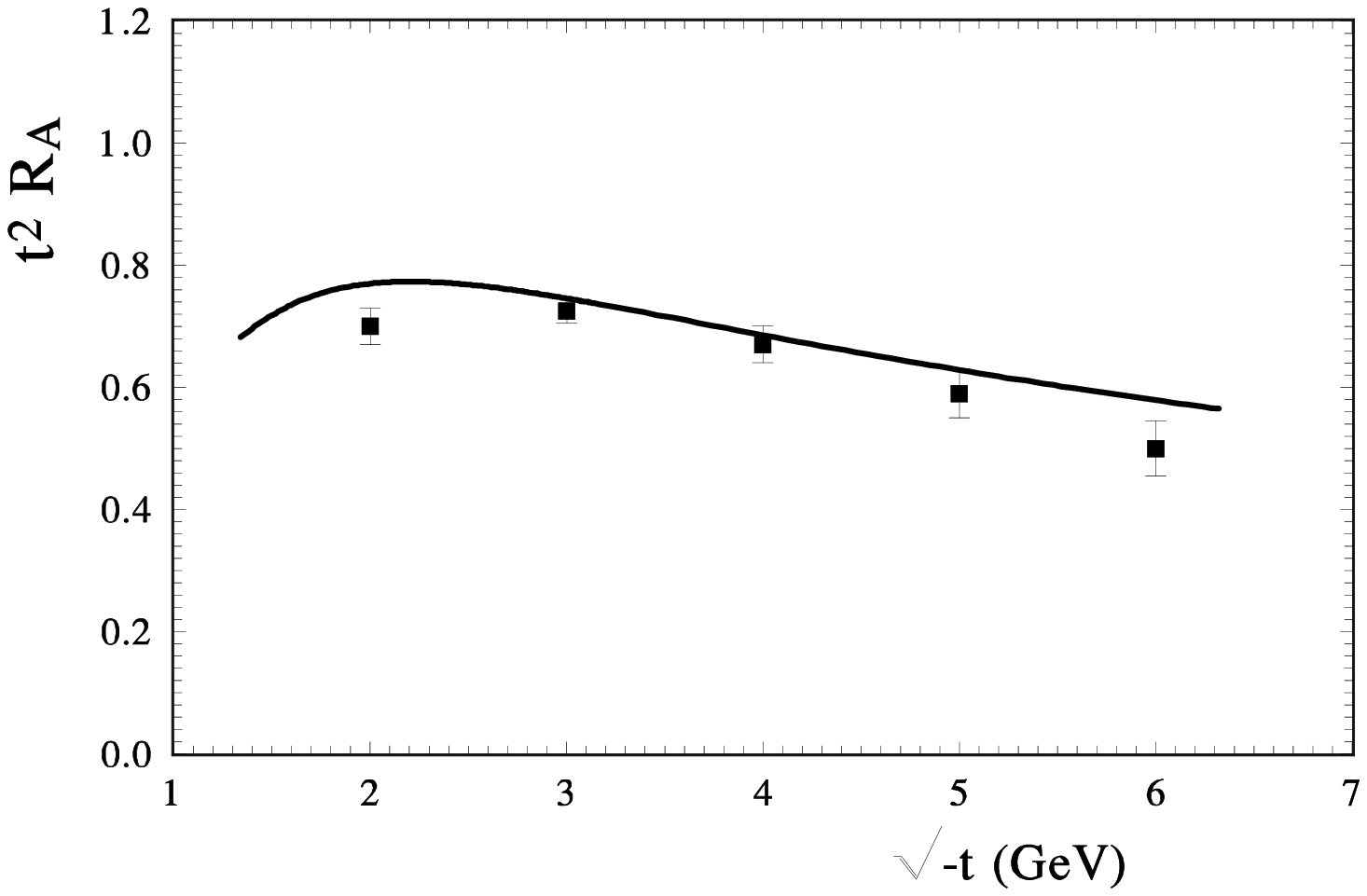} %{ffdm.ps}
\includegraphics[width=.3\textwidth]{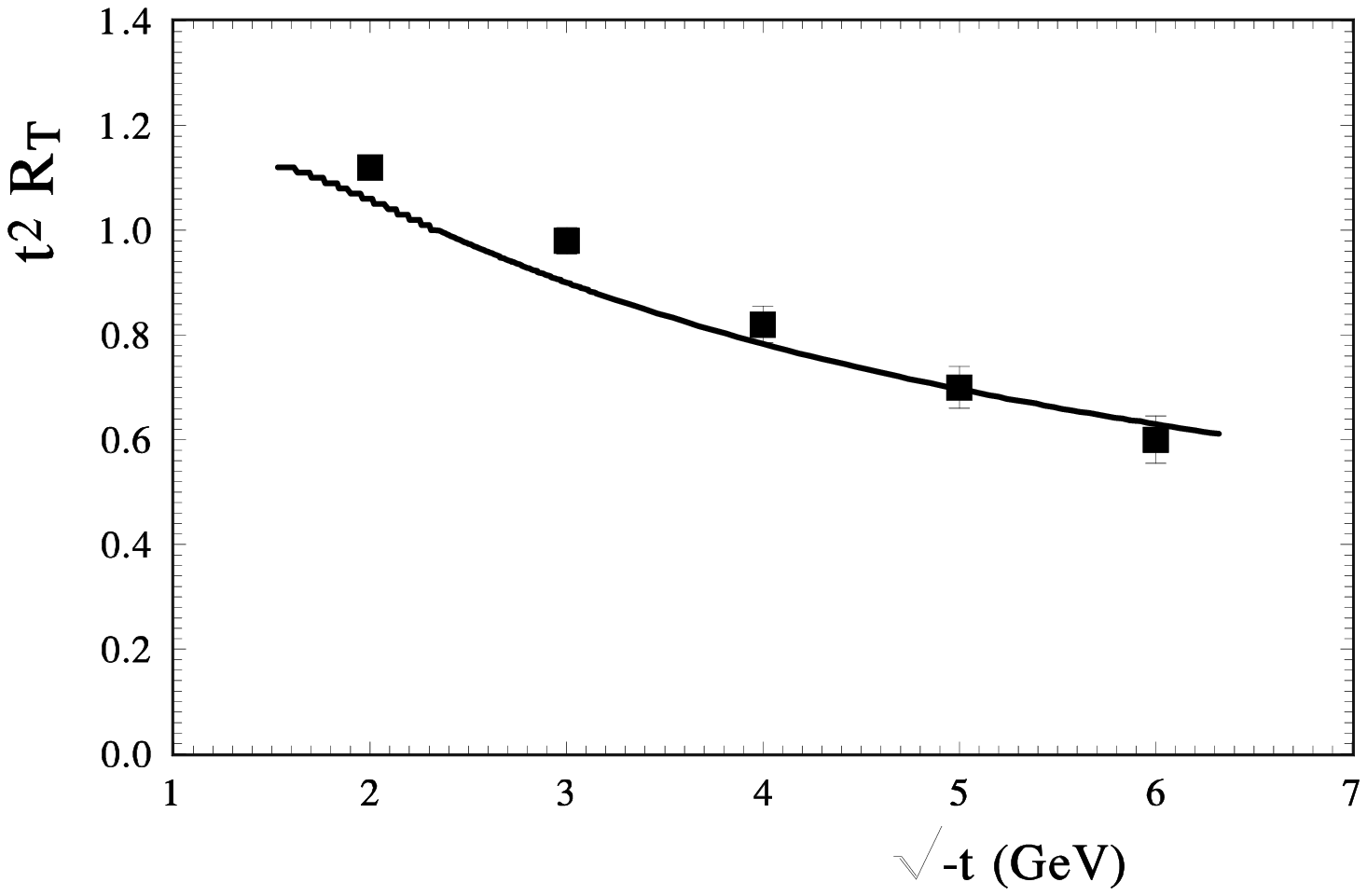} %{ffdm.ps}
\caption{ Compton form factors (squares - show the model calculations \cite{DK-13} \\
 a) [left] $t^2 R_{V}(t)$ , b) [middle] $t^2 R_{T}(t)$  , c) [right]  $t^2 R_{A}(t)$  }.
\label{Fig_14}
\end{figure}

%\vspace{1cm}
% \vspace{-1cm}
%%%Fig 14
\begin{figure}
\includegraphics[width=.45\textwidth]{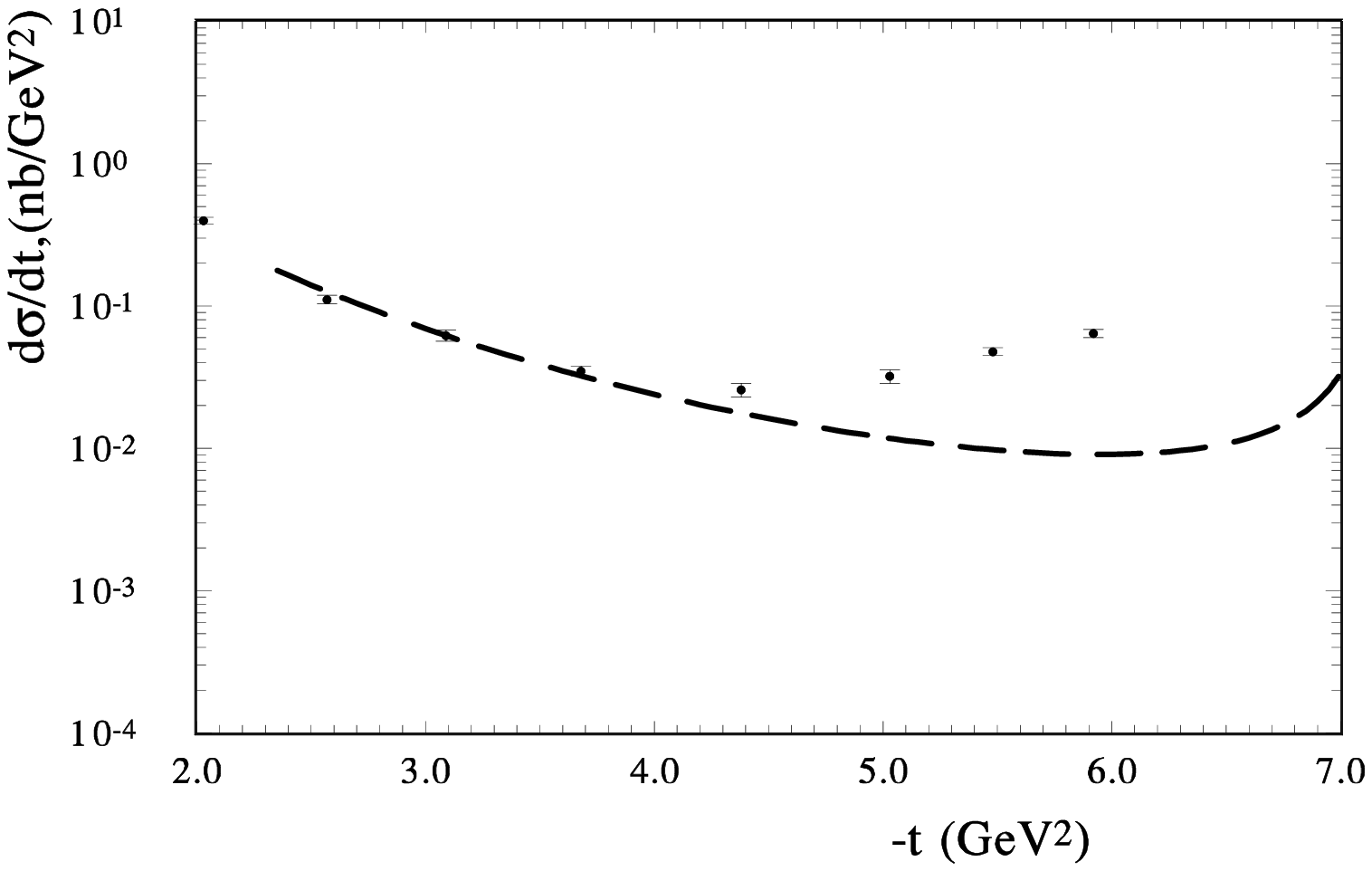} %{ffdm.ps}
\includegraphics[width=.45\textwidth]{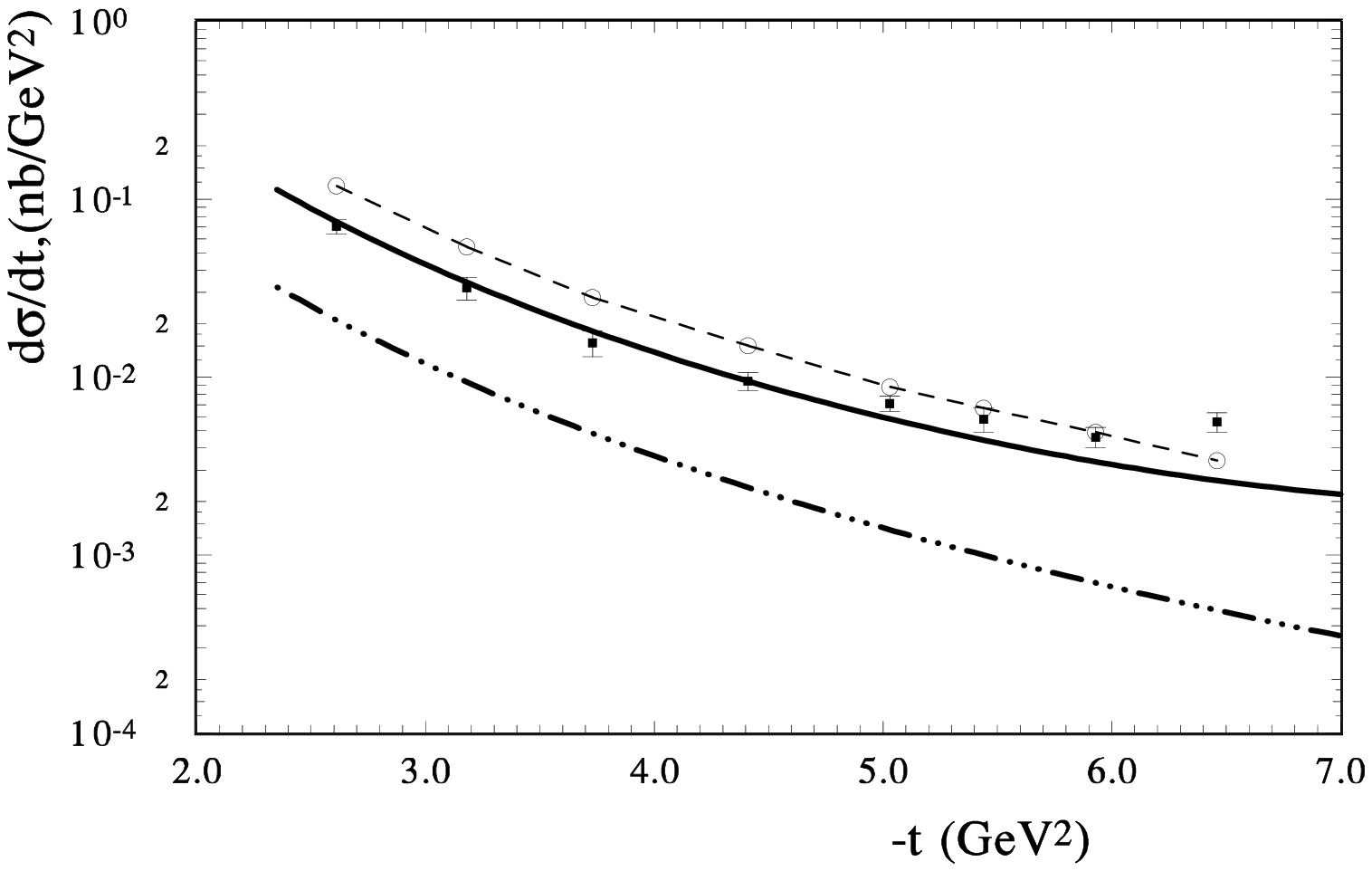} %{ffdm.ps}
\caption{
  Differential Compton cross sections $\gamma p \rightarrow \gamma p$;
 the curves are our calculations at \\
 a) [left] $s=8.9$ GeV$^2$, \
  b) [right] $s=10.92$ GeV$^2$ (dashed line with circles - the calculations  \cite{DK-13}),
 and $s=20$ GeV$^2$
 % ( hard line, dashed line, dot-dashed line with (x), med-dashed line with ($+$),
 %  correspond to the PDFs [50,52,53,48];
   the data points are for $s=8.9$ GeV$^2$ (circles);
  $s=10.92$ GeV$^2$ (squares) \cite{105-Dan07}.
  }
\label{Fig_14}
\end{figure}
%\vspace{1cm}
% \vspace{-1cm}

     The calculations of  the differential cross sections of RCS are shown  in Fig.15 at
     three energies $s = 9.8, 10.92$ and $20$ GeV$^2$. The calculations have a sufficiently
     good coincidence with the existing experimental data and, in some part, coincide with
     calculations \cite{DK-13}.

\section{Elastic nucleon scattering (HEGS model) }

  Let us use the obtained momentum transfer dependence of GPDs in calculations of the electromagnetis and gravimagnetic form factors
  of the nucleons as the first and second moments of GPDs.
  The obtained form factors are related to the charge and matter distributions.
%  Fixing the parameters of the obtained form factors
%  the new High Energy General Structure (HEGS) model of the elastic proton-proton and proton-antiproton scattering
%  was proposed \cite{HEGS0,GPD-PRD14}.
  The new High Energy General Structure (HEGS) model of the elastic proton-proton and proton-antiproton scattering
  was proposed \cite{HEGS0,HEGS1}   with the two form factors determined early with  the fixed parameters.
% \cite{HEGS0,GPD-PRD14}

  The model is very simple from the viewpoint of the number of parameters and functions.
  There are no any artificial functions or any cuts which bound the separate
  parts of the amplitude by some region of momentum transfer or energy.
  One of the most remarkable properties is that the real part of the
  hadron scattering amplitude is determined only by complex energy $\hat{s}$
  that satisfies the crossing-symmetries.

         The differential cross
  sections of nucleon-nucleon elastic scattering  can be written as the sum of different
  helicity  amplitudes:
\begin{eqnarray}
  \frac{d\sigma}{dt} =
 \frac{2 \pi}{s^{2}} (|\Phi_{1}|^{2} +|\Phi_{2}|^{2} +|\Phi_{3}|^{2}
  +|\Phi_{4}|^{2}
  +4 | \Phi_{5}|^{2} ). \label{dsdt}
\end{eqnarray}

%!!!!\renewcommand{\bottomfraction}{0.7}
%

  The total helicity amplitudes can be written as $\Phi_{i}(s,t) =
  F^{h}_{i}(s,t)+F^{\rm em}_{i}(s,t) e^{\varphi(s,t)} $\,, where
 $F^{h}_{i}(s,t) $ comes from the strong interactions,
 $F^{\rm em}_{i}(s,t) $ from the electromagnetic interactions and
 $\varphi(s,t) $
 is the interference phase factor between the electromagnetic and strong
 interactions \cite{selmp1,selmp2,Selphase}.
% For the hadron part the amplitude with spin-flip is neglected in this approximation, as usual at high % energy.

 The  (HEGS)  model \cite{HEGS1}  gives a quantitative     description of the elastic nucleon scattering
 at high energy  with only $5$ fitting high energy parameters.
  A successful description  of the existing experimental data by the model shows that
   the elastic scattering  is determined by the generalized structure of the hadron.
The model leads to a  coincidence of the model calculations with the preliminary data at 8 TeV.
   We found that the  standard eikonal approximation \cite{Unit-PRD}   works perfectly well from
  $\sqrt{s}=9$  GeV up to  8 TeV.

  In the model the Born term of the elastic hadron amplitude is determined
  \begin{eqnarray}%\ba
 F_{h}^{Born}(s,t) \ =  h_1 \ F_{1}^{2}(t) \ F_{a}(s,t) \ (1+r_1/\hat{s}^{0.5}) % \\ \nonumber
    \    +  h_{2} \  A^{2}(t) \ F_{b}(s,t) \ (1+r_2/\hat{s}^{0.5}),
    \label{FB}
% \nonumber
\end{eqnarray}
  where $F_{a}(s,t)$ and $F_{b}(s,t)$  have the standard Regge form % correspond
  \begin{eqnarray}%\ba
 F_{a}(s,t) \ = \hat{s}^{\epsilon_1} \ e^{B(s) \ t}; \ \ \
 F_{b}(s,t) \ = \hat{s}^{\epsilon_1} \ e^{B(s)/4 \ t}.
\label{FB-ab}
% \nonumber
\end{eqnarray}

  The final elastic  hadron scattering amplitude is obtained after unitarization of the  Born term.
    So, first, we have to calculate the eikonal phase and then, using the standard eikonal representation,
    obtained the final hadron elastic scattering amplitude.

     Our complete fit of 3416 experimental data in the energy range
      $9.8 \leq \sqrt{s} \leq 8000 \ $ GeV
      and the region  of the momentum transfer
      $0.000375 \leq \ -t \ \leq 14.75 \ $GeV$^2$ gave
 % $N_{exp.}=3296;  \ \ \
  $ \sum_{i=1}^{N} \chi_{i}^{2}/N=1.28$
with the parameters
$h_1=3.67; \ \ h_2=1.39; \ \  h_{odd}=0.76; \ \
 k_0=0.16; \ \  r_{0}^{2}=3.82; \ \ \ $ and the low energy parameters
 $ \ h_{sf} = 0.05; \ \ r_1=53.7; \ \ \ \ r_2 = 4.45$.

% \begin{multicols}{2}
% \end{multicols}
%  \begin{twocolumn}
%\end{twocolumn}

           Obviously, for such a huge energy region we have a very small
   number of free parameters.
  Noteworthy also is a good description of the CNI region of momentum transfer
  in a very wide energy region,
   approximately three orders, with the same slope of the scattering amplitude. % (for example, Fig6a).
  As example, the differential cross sections % of the elastic scattering % $pp$ and $p\bar{p}$ at
 of the proton-proton elastic scattering % and proton-antiproton
  in the region of the diffraction minimum and different energies
 are presented   % in  Fig.3  %at   $\sqrt{s}= 9.8 $ GeV   for $pp$ scattering,
%   and $\sqrt{s}= 11 $ GeV for  $p\bar{p}$ elastic scattering (Fig.5a),
 in Fig. 3 at $\sqrt{s}= 30.5 $ GeV  and $\sqrt{s}= 7. $ TeV. % for $pp$ scattering (Fig.3).
   In most part, the form and energy dependence of the diffraction  region is determined by the real part of the
   scattering amplitude. Note that in the model only the Born term is determined. The complicated diffractive
   structure is obtained only after the unitarization procedure.
   The extended variant of the model \cite{GPD-PRD14} shows the contribution of the "maximal" odderon with specific
   kinematic properties
  and does not show a visible contribution of the hard pomeron, as in \cite{NP-HP}.
%   The model quantitatively reproduces the differential cross sections in the whole examined energy region
% in spite of the fact that the size of the slope  is changing essentially in this region
% (due to the standard Regge behavior $log(\hat{s})$)
%  and the real part of the scattering amplitude has a different behavior for
%   $pp$ and $p\bar{p}$.

%	left text

%begin{table}[ht]
%  \centering
%  \begin{tabular}{|p{3cm}|c|p{3cm}|}
%\vspace{1cm}
% \vspace{-1cm}
%%%Fig 14
\begin{figure}
%\begin{minipage}{0.5\textwidth}
%  \begin{flushleft}
%\vspace{3cm}
%\begin{center}
%\mbox{\epsfysize=60mm\epsffile{GedGd.eps}}x
\includegraphics[width=.3\textwidth]{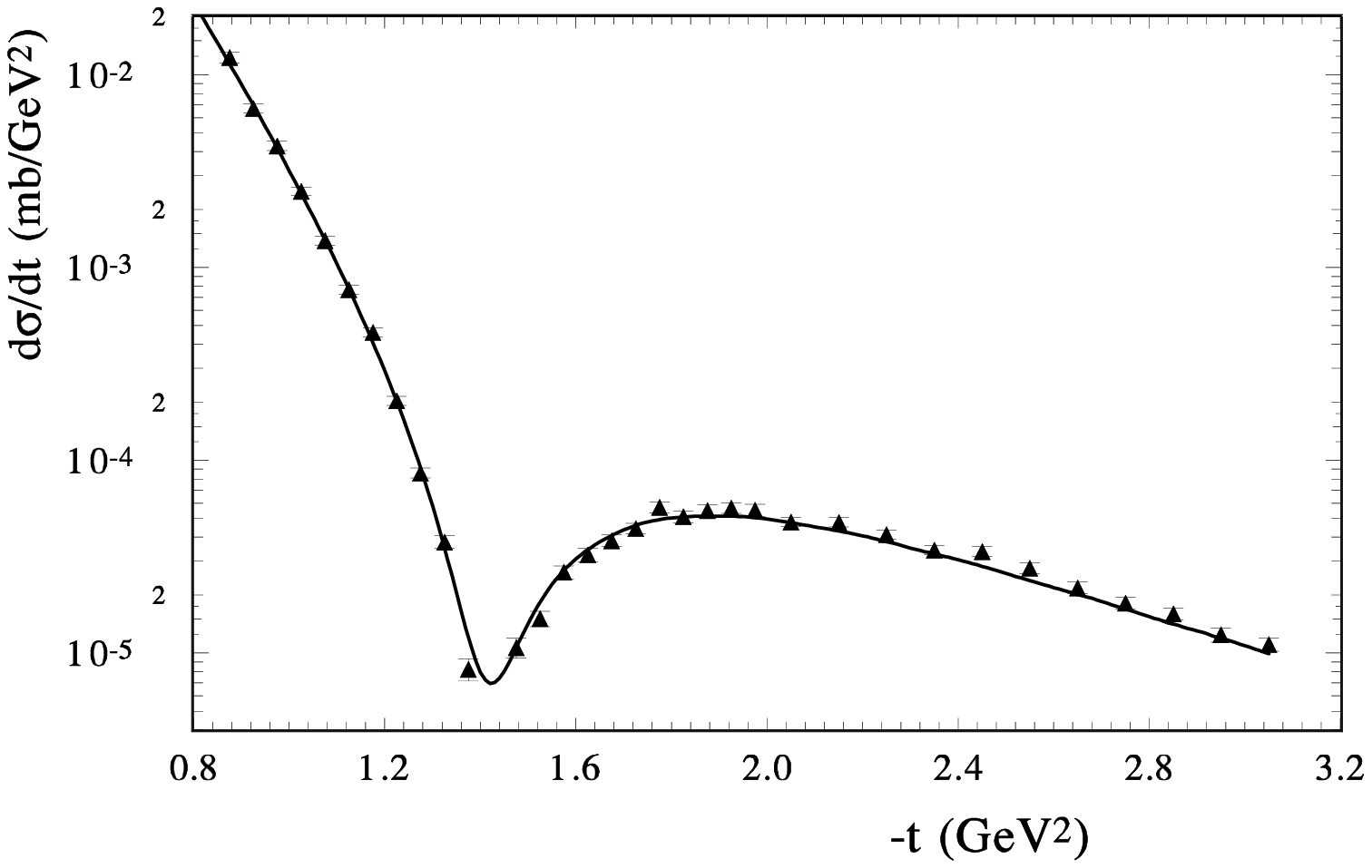} %{ffdm.ps}
\includegraphics[width=.3\textwidth]{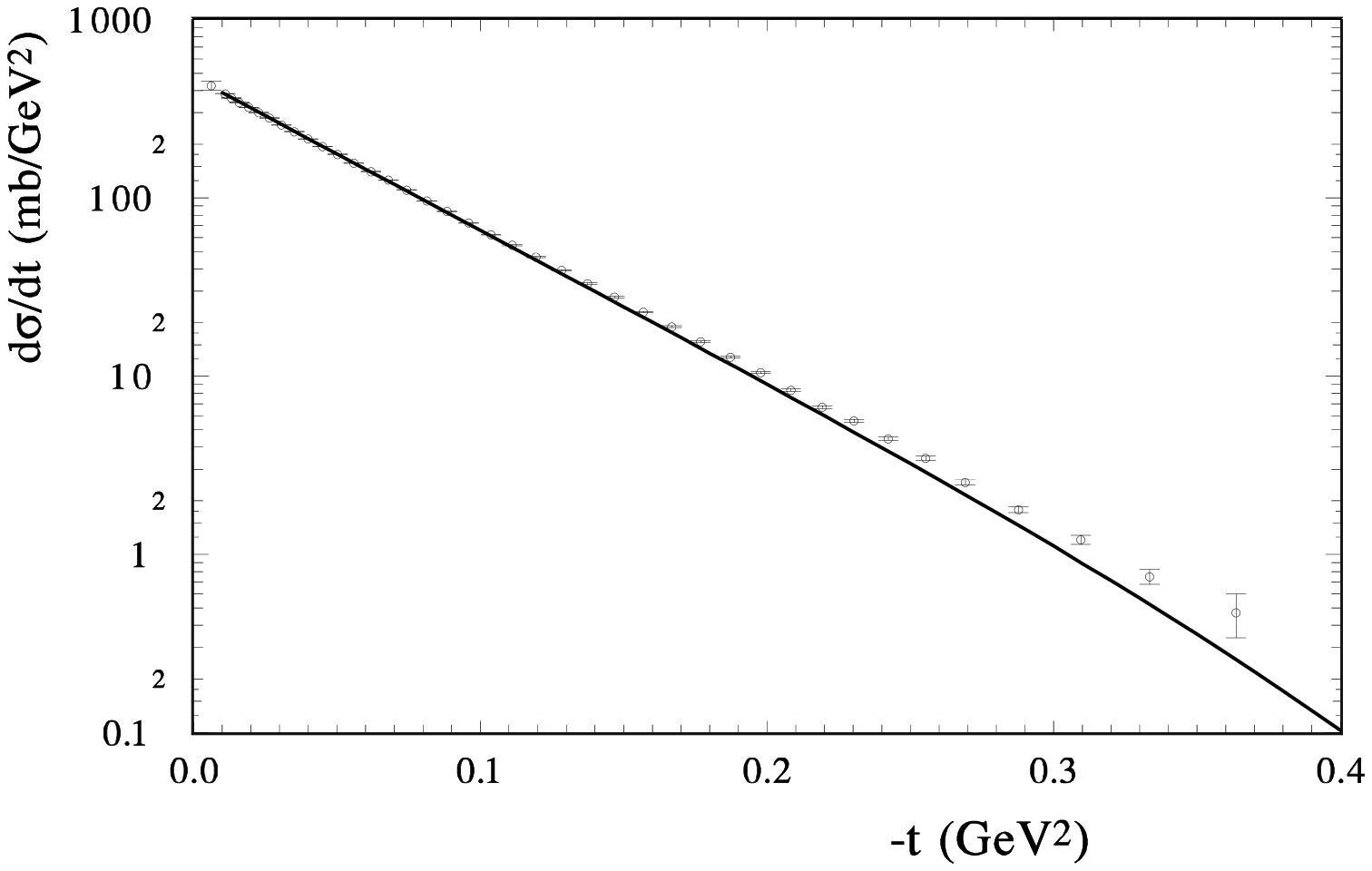} %{ffdm.ps}
\includegraphics[width=.3\textwidth]{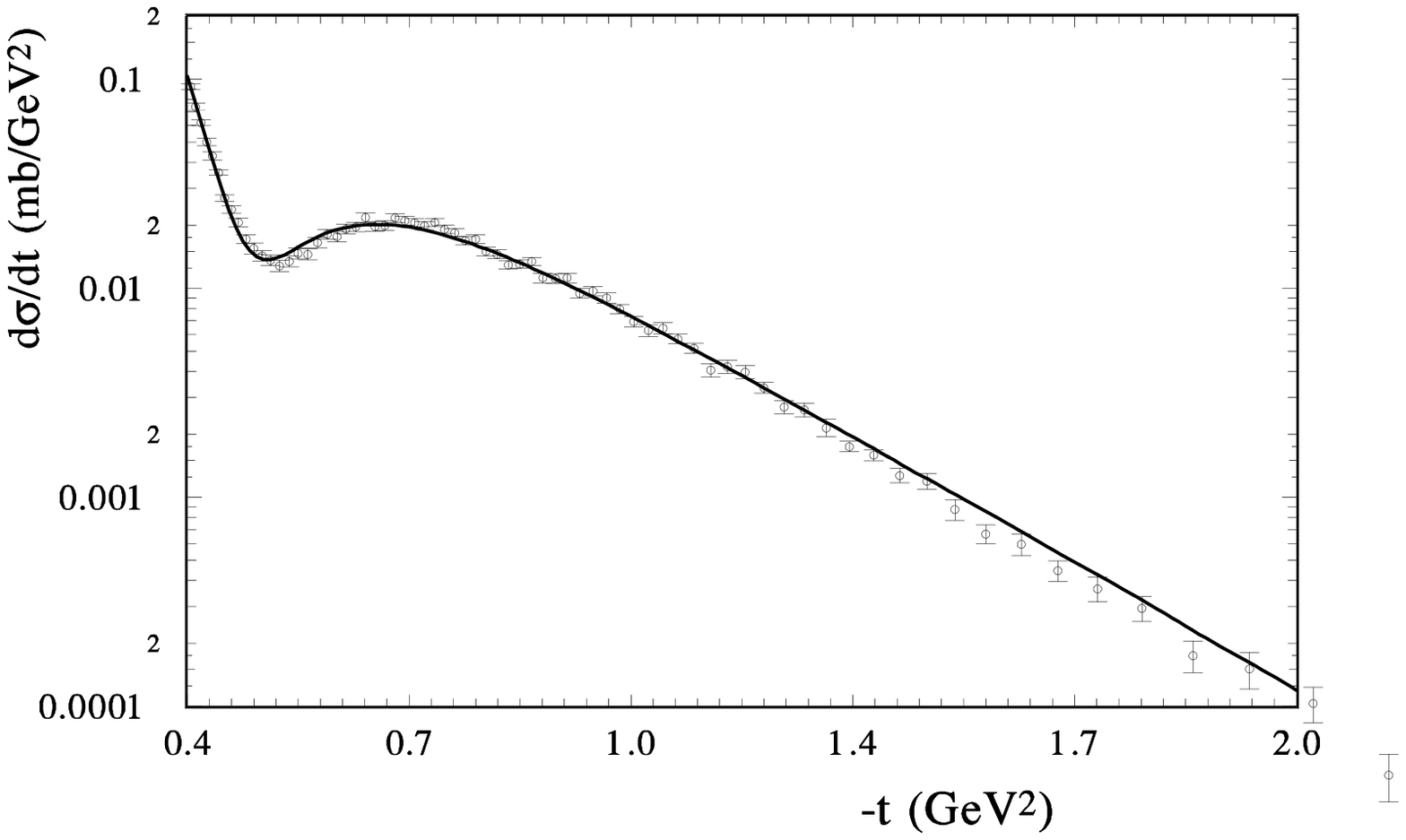}
\caption{  Differential cross sections of the elastic $pp$ scattering
a) [left] at $\sqrt{s}=30.5$ GeV
%in the region of the diffraction minimum
b)[middle] at $\sqrt{s}= 7$ TeV (small $t$) and c) [right] at $\sqrt{s}= 7$ TeV (large $t$)
  }
\label{Fig_14}
\end{figure}

  Only our model takes into account the whole region of the momentum transfer
  ($3.75 \ 10^{-4} \geq |t| \geq 15$ GeV$^{2}$, which includes the high precision experimental data
  in the Coulomb-hadron interference region.

\section{Conclusion }
 The analysis of PDFs  of  many Collaborations with different forms of the $x$-dependence
    show that additional parameters of the function $f(x)$ in the exponential  $t$-dependence part of GPDs
    lead to  the
   identical results in the final fitting procedure. % in the descriptions of the electromagnetic form factors.
   Hence, such additional parameters  equalize the different form of PDFs to obtain the same results for the
   electromagnetic form factors of the nucleons. % The analysis of PDFs  of the many Collaborations with different form of the $x$-dependence

       Our examination of the obtained momentum transfer dependence of GPDs was checked in the case of the real Compton scattering.
  The calculated  Compton form factors $R_{V}(t)$,   $R_{T}(t)$, $R_{A}(t)$ gave a somewhat larger slope than
  such form factors calculated in \cite{DK-13}  (see Fig.1). Our form factors reproduce the differential cross sections of the real Compton
  scattering at large momentum transfer reasonably well (see Fig.2).

 %  The first and second moments of same GPDs give the electromagnetic and gravimagnetic form factors.
 %
    GPDs let us calculate not only electromagnetic form factor (which reflects the charge distribution) but also the
    second moment of the GPDs give the gravimagnetic form factors (which  reflects the matter distribution of the nucleons).
 The new HEGS model  of the elastic % proton-proton and proton-antiproton scattering
 $pp$ and $\bar{p}p$ scattering  was built taking into account both these form factors.
    It gave the possibility  to essentially decrease the number of the fitting parameters.
    As a result, the model with a minimum of the fitting parameters
    describes quantitatively  the maximum number of the experimental data in a wide region of the energy ($9.8$ GeV $\leq \sqrt{s} \leq 8$ TeV)
    and the momentum transfer ($0.00037 \leq |t| \leq 15$ GeV$^2$). Such a unique description of the differential cross sections
    on the basis of the  GPDs gave large support of the determining momentum transfer dependence of the GPDs.

\vspace{0.5cm}
{\bf Acknowledgments}
%\section{Acknowledgments}
 {\it The authors would like to thank J.-R. Cudell and O.V. Teryaev
   for fruitful   discussion of some questions   considered in the paper.} \\

%\begin{footnotesize}
%\bibliography{abbr_long,pubext}

% {\footnotesize
%\begin{spacing}{0.5}
%{\small

\end{document}